\documentclass[prl,superscriptaddress,floatfix,showpacs,aps,reprint,letterpaper,nobalancelastpage]{revtex4-1}
\usepackage{graphicx} 
\usepackage{color} 
\usepackage{amsmath} 
\usepackage{hyperref} 
\usepackage{amssymb}
\newcommand{\ket}[1]{\left\lvert #1 \right\rangle}

\newcommand{\MHz}{\mathrm{MHz}}
\newcommand{\GHz}{\mathrm{GHz}}
\newcommand{\us}{\mu\mathrm{s}}
\newcommand{\ns}{\mathrm{ns}}

\newcommand{\wcav}{\omega_{\mathrm{C}}}

\newcommand{\bra}[1]{\langle{#1}|}

\begin{document}
\title{Implementing optimal control pulse shaping for improved single-qubit gates
}
\date{May 7, 2010}
\author{J. M. Chow}
\author{L. DiCarlo}
\affiliation{Departments of Physics and Applied Physics, Yale University, New Haven, Connecticut 06520, USA}
\author{J. M. Gambetta}
\author{F. Motzoi}
\affiliation{Institute for Quantum Computing and Department of Physics and Astronomy, University of Waterloo, Waterloo, Ontario, Canada N2L 3G1}
\author{L. Frunzio}
\author{S. M. Girvin}
\author{R. J. Schoelkopf}
\affiliation{Departments of Physics and Applied Physics, Yale University, New Haven, Connecticut 06520, USA}

\begin{abstract}
We employ pulse shaping to abate single-qubit gate errors arising from the weak anharmonicity of transmon superconducting qubits. By applying shaped pulses to both quadratures of rotation, a phase error induced by the presence of higher levels is corrected. Using a derivative of the control on the quadrature channel, we are able to remove the effect of the anharmonic levels for multiple qubits coupled to a microwave resonator.  Randomized benchmarking is used to quantify the average error per gate, achieving a minimum of $0.007\pm0.005$ using $4\,\ns$-wide pulse.
\end{abstract}
\maketitle

The successful realization of quantum information processing hinges upon the ability to perform high-fidelity control (gates) of quantum bits, or qubits. A value of $10^{-3}$ error per gate (EPG) is typically quoted as the necessary threshold for fault-tolerant quantum computation \cite{knill_nature_2005,*al:2007nx}. For any two-level system, the optimal achievable gate performance is set by the ratio of gate time to coherence time. However, quantum information processing systems consist of multiple coupled qubits. Often these qubits are not truly two-level systems, but rather quantum objects with a rich level structure. Thus, one important challenge is to achieve optimized single-qubit gates in a large Hilbert space.

Optimal control theory has been previously employed in nuclear magnetic resonance, non-linear optics, and trapped ions to combat specific errors such as always-on qubit couplings and spatial inhomogeneities \cite{Khaneja:2005kx}. In these systems, even with optimized qubit control, limits to qubit gates are often due to systematic errors rather than decoherence. Superconducting qubit control, however, has primarily been limited by coherence times. With recent progress such as the demonstration of high-fidelity single-qubit gates \cite{lucero_gates_2008,chow_bm_2009}, two-qubit gates \cite{Yamamoto:2003bv,Plantenberg:2007it,dicarlo_2009,bialczak-2009}, entanglement~\cite{steffen_entang_2006,Ansmann:2009yq,chow_entang_2009} and simple quantum algorithms \cite{dicarlo_2009}, the superconducting qubit architecture is growing into a more complex quantum information testbed, placing the level of qubit control under increased scrutiny. One approach towards improving single-qubit gates is to decrease the total gate time. However, in multi-level qubits such as the transmon \cite{koch_charge-insensitive_2007} or phase qubit~\cite{lucero_gates_2008}, the weak anharmonicity sets a lower limit on the gate time. Furthermore, superconducting qubit coupling schemes such as circuit quantum electrodynamics (QED), in which a transmission-line cavity couples multiple qubits~\cite{majer_coupling_2007,Sillanpaa_2007}, can make single-qubit control more difficult as a result of many extra levels in the Hilbert space.

In this Letter, we implement simple optimal control techniques to improve single-qubit gates in a circuit QED device with two superconducting transmons. The pulse-shaping protocols we investigate are guided by the recent theoretical exploration of derivative removal via adiabatic gate (DRAG) in Ref.\,\cite{Motzoi:2009ca}. We demonstrate the improvement of single-qubit gates on two separate qubits in the same device using the first-order correction in DRAG, by switching from rotations around a single axis induced by Gaussian-modulated microwave tones to rotations performed using Gaussians and derivatives of Gaussians applied on two perpendicular axes. We tune up the pulses using a set of calibration rotation experiments with results that agree with the model outlined in Ref.\,\cite{Motzoi:2009ca}.  Randomized benchmarking (RB) is employed to show the reduction of the average single-qubit gate error~\cite{knill_randomized_2008,ryan_randomized_2008,chow_bm_2009}. For the shortest gate width of 4\,ns, we find an improvement by a factor of $\sim 15$ down to a minimum EPG of $0.007\pm 0.005$, which is at the limit imposed by two-level relaxation. Independently, the DRAG technique is also being implemented at UCSB with superconducting phase qubits \cite{lucero_inprep}. 

The optimal control technique of DRAG in Ref.\,\cite{Motzoi:2009ca} prescribes a simple pulse-shaping protocol for reducing single-qubit gate errors due to the presence of a third level. Neglecting the cavity, which is detuned away from any transitions, the driven three-level system, or qutrit, is described by the Hamiltonian
\begin{equation}
	H= \hbar \sum_{j=1,2}\left[\omega_{j-1,j} \ket{j}\bra{j}+\mathcal{E}(t)\lambda_j \left(\sigma_j^+ +\sigma_j^-\right)\right],
	\label{eq:hamiltonian}
\end{equation}
where $\sigma_j^- = \ket{j-1}\bra{j}$ and $\sigma_j^+ = \ket{j}\bra{j-1}$ are lowering and raising operators, $\omega_{j,j-1}$ are transition energies with the ground state energy set to zero, and $\lambda_1 =1$, $\lambda_2= \lambda=\Omega_{1,2}/\Omega_{0,1}$, give the relative drive coupling strengths of the $0\leftrightarrow1$ and $1\leftrightarrow2$ transitions, $\Omega_{0,1}$ and $\Omega_{1,2}$, respectively. We can define the anharmonicity of the system as the difference between the $0\leftrightarrow1$ and $1\leftrightarrow2$ transition frequencies, $\alpha_1= \omega_{1,2}-\omega_{0,1}$. The drive is only turned on for a fixed gate duration $t_g$ and given by $\mathcal{E}(t) = \mathcal{E}^x(t)\cos(\omega_{\text{d}} t)+ \mathcal{E}^y(t)\sin(\omega_{\text{d}}t)$,
where $\mathcal{E}^{x,y}$ are independent quadrature controls. 

Given a large $\alpha_1\gg\omega_{0,1}$, the effective Rabi drive rate $\Omega_{1,2}$ to induce any direct or time-dependent transitions to $\ket{2}$ can be negligible. However, for a system such as the transmon, $\alpha_1$ is only $\sim 3-5\%$ of $\omega_{0,1}$. There are two specific qubit gate errors which arise due to this reduced anharmonicity. With a non-negligible $\Omega_{1,2}$, it becomes possible for the gate targeting the $0\leftrightarrow1$ transition to directly populate $\ket{2}$, leaving the qubit subspace. However, a second and more dominant error is a temporary population of $\ket{2}$ during the course of a control pulse to the $0\leftrightarrow 1$ transition, leading to the addition of a phase rotation to the intended gate. Although Gaussian control pulses (characterized by a width $\sigma$) are often the paradigm due to their localized frequency bandwidths given by $B= 1/2\pi\sigma$, leakage errors can occur as gate times are reduced such that $B$ is comparable to $\alpha_1$. A simplified correction protocol to the leakage errors as prescribed in Ref.\,\cite{Motzoi:2009ca} is to apply an additional control on the quadrature channel, $\mathcal{E}^y(t) = \beta  \dot{\mathcal{E}}^x(t)$ and a dynamical detuning of the drive frequency $\delta(t) =  \mathcal{E}^x(t)^2(-4 \beta \alpha_1+\lambda^2)/4\alpha_1$, where $\beta$ is a scale parameter. For a qutrit driven without dynamical detuning \cite{Motzoi:inprep}, the optimal $\beta=\lambda^2/4\alpha_1$.

\begin{figure}[t!]
\centering
\includegraphics[scale=1.0]{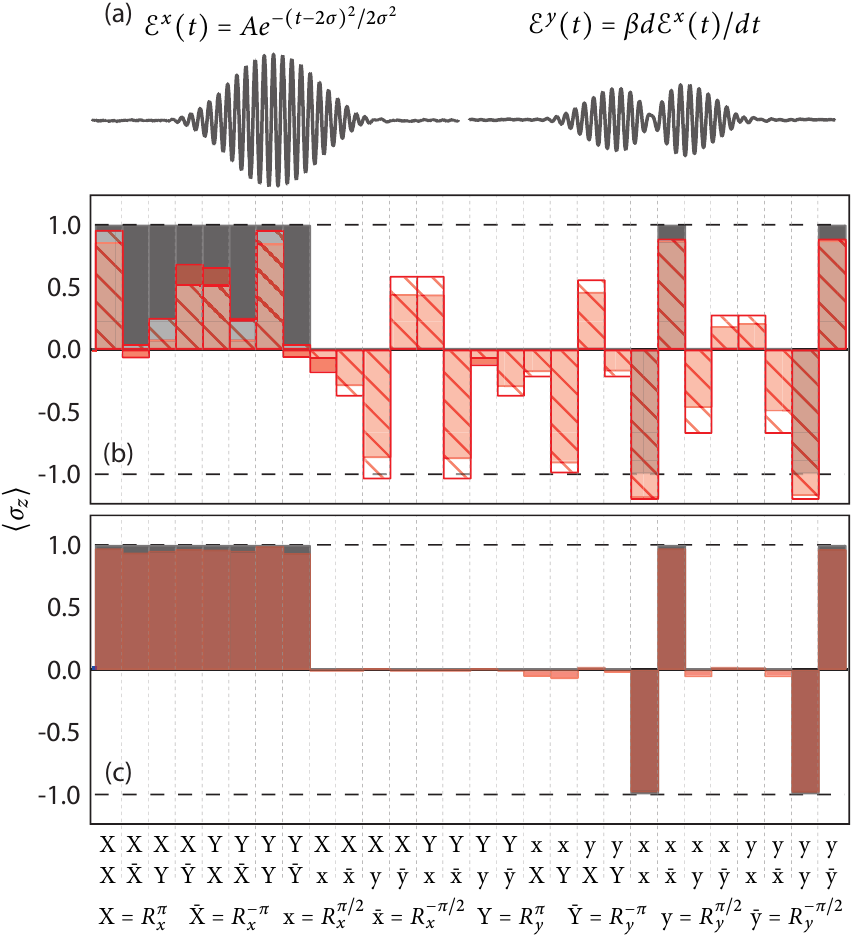}
\caption{(color online)~(a)~Gaussian and derivative pulse shapes applied to the in-phase and quadrature control channels, respectively, for implementing DRAG to first order. (b)~Measured $\langle \sigma_z\rangle$ on qubit L (shaded red bars) with Gaussians with $\sigma=1\,\ns$  for a test sequence consisting of pairs of $\pi$ and $\pi/2$ rotations. The slash filled bars correspond to a master-equation simulation of a three-level system with parameters of the sample tested. The grey filled bars reflect ideal values. (c)~Similarly measured $\langle \sigma_z\rangle$ on qubit L but using derivative pulses on the quadrature with a scale factor $\beta^{\text{L}}=0.4$ (shaded red bars), also overlaid on the ideal values (shaded dark grey bars).}
\label{fig:fig1}
\end{figure}

The experiments are performed in a circuit QED sample consisting of two transmons coupled to a coplanar waveguide resonator. The sample fabrication and experimental setup are described in Ref.\,\cite{dicarlo_2009}. The two transmons (designated L and R) are detuned from one another with ground to excited state ($0\leftrightarrow 1$) transition frequencies of $\omega_{0,1}^{\text{L,R}}/2\pi=8.210,\, 9.645\,\GHz$ and the ground state cavity frequency is $\wcav/2\pi=6.902\,\GHz$. The anharmonicities of the transmons are found using two-tone spectroscopy measurements \cite{schreier_suppressing_2008} to be  $\alpha_1^{\text{L,R}} = 330,\,300\,\MHz$ and coherence times are measured to be $T_1^{\text{L,R}}=1.2,\,0.9\,\us$ and $T_2^{*\text{L,R}} = 1.5,\,1.1\,\us$.

To implement DRAG to first order and perform single-qubit gates on the transmons, we use an arbitrary-waveform generator to shape microwave-frequency pulses with quadrature control, permitting rotations about either the $x$- or $y$-axis of each qubit. We fix the drive frequency to $\omega_{0,1}$, and the pulse amplitudes and phases define the rotation angle and axis orientation, respectively. When performing an $x$ rotation, $\mathcal{E}^x(t)$ is a Gaussian pulse shape [Fig.\,\ref{fig:fig1}(a)], while the derivative of the Gaussian is applied simultaneously on the other quadrature, $\mathcal{E}^y(t)=\beta \dot{\mathcal{E}}^x(t)$. All of the pulses are truncated to $2\sigma$ from the center with an added buffer time of 5\,ns to ensure complete separation between concatenated pulses.

A simple test sequence is used to tune up the scale parameter $\beta$ as well as to demonstrate the effect of using first-order DRAG pulses versus standard single-quadrature Gaussians. The sequence consists of pairs of $\pi$ and $\pi/2$ pulses around both the $x$- and $y$-axes. An important feature of this sequence is that the final average $z$-projection of the single qubit, $\langle \sigma_z \rangle$, will ideally take on values from the set $S=\{+1,0,-1\}$, making any deviations easily visible. 

Using Gaussian pulse shaping with $\sigma=1\,\ns$ and implementing the test sequence for qubit L, we find significant deviations from $S$, as shown in the solid red bars of Fig.\,\ref{fig:fig1}(b). The theoretical results for each pair of rotations are shown with solid grey bars in the background. Note that there are some concatenated rotations for which $|\langle \sigma_z \rangle| > 1$, which is because the measurement calibration is performed using a $\pi$ pulse which is itself non-ideal and plagued by the same errors induced by the third level. We observe that the largest errors occur when the two rotations are around different axes, which indicates the presence of significant phase error, or a residual $z$ rotation after the first gate. 

We repeat the same test sequence, but applying the derivative of the Gaussian  to the quadrature channel. By varying $\beta$, it is possible to find an optimal value such that the measurements of $\langle\sigma_z\rangle$ agree very well with the theoretical predictions. The shaded red bars of Fig.\,\ref{fig:fig1}(c) show measured $\langle \sigma_z \rangle$ for qubit L using $\beta=0.4$. Here, deviations from the ideal grey bars decrease to $<2\%$. We have also applied the DRAG protocol for qubit R, finding the  optimal value $\beta=0.25$ (data not shown). From the experimental determination of $\beta$ and $\alpha_1$, we can infer the second excited state coupling strengths $\lambda^{\text{L,R}} = 1.82,\,1.41$. Using $\lambda^{\text{L}}$ and the three-level model of Eq.\,\ref{eq:hamiltonian}, a master equation simulation for the Gaussian shaping gives the red hash-filled bars in Fig.\,\ref{fig:fig1}(a), which demonstrates good agreement with the experiment.

We find excellent agreement for $\lambda$ with a simple calculation for the anticipated $\lambda$ in a cavity-transmon coupled system. The cavity modifies the drive strengths $\Omega_{0,1}$ and $\Omega_{1,2}$ due to its filtering effect. Specifically, for a transmon in a cavity, we have
\begin{equation}
	\Omega_{j-1,j} = \mathcal{E} \sum_n\frac{\gamma^n\sqrt{n}g_{j-1,j}}{n\wcav-\omega_{j-1,j}}, 
	\label{eq:drive}
\end{equation} where $n$ indexes the cavity mode, $j=1,\,2$ for the transmon excitation level, $\gamma=\{1,-1\}$ depending on whether the qubit is located at the input or output side of the cavity, and $g_{i,j}$ is the matrix element coupling the $i\leftrightarrow j$ transmon transition to the cavity \cite{koch_charge-insensitive_2007}. Using the relevant parameters of the two transmons in the experiment and including only the fundamental mode of the cavity, we find $\lambda^{\text{L,R)}} = 1.85,\,1.57$,  within $12\%$ of those determined from the test sequence. There are corrections to the drive due to the higher modes of the cavity, but it is difficult to use Eq.\,(\ref{eq:drive}) to estimate as a result of cutoff dependence.

\begin{figure}[tb!]
\centering
\includegraphics[scale=1.0]{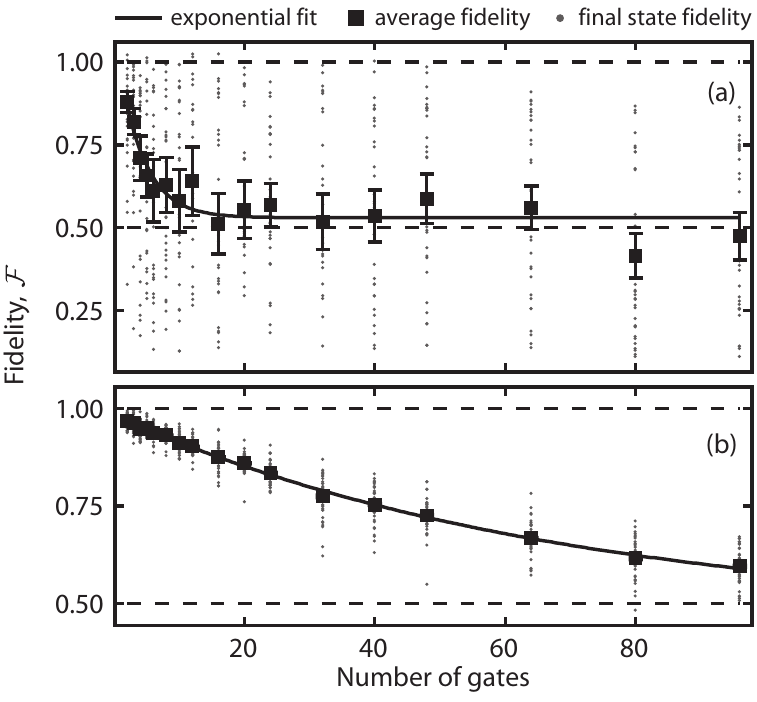}
\caption{Randomized benchmarking for qubit L with $\sigma = 1\,\ns$ using (a)  Gaussian pulses, and (b) additional Gaussian derivative pulses on the quadrature channel. The scattered grey points are extracted fidelities for 32 RB sequences, truncated at different numbers of gates. A remarkable reduction in the extracted average error per gate (black squares) of the benchmarking results is observed going from (a) to (b). The error bars indicate the variance of all the RB sequences and are smaller than the squares in (b).}
\label{fig:fig2}
\end{figure}

We characterize the degree of improvement to single-qubit gates by using the technique of randomized benchmarking (RB)~\cite{knill_randomized_2008}. RB allows us to determine the average error per gate through the application of long sequences of alternating Clifford gates ($R_{x,y}^{\pi/2}$) and Pauli gates, chosen from $\{\openone, R_x^{\pi}, R_y^{\pi}, R_z^{\pi}\}$ \cite{gottesman-1998}. We use the RB pulse sequences originally given in Ref.\,\cite{knill_randomized_2008} and adapted to superconducting qubits in Ref.\,\cite{chow_bm_2009} for both the Gaussian and the derivative pulse shaping for transmon L. We truncate the randomized sequences at various lengths and compare the resulting measurement of $\langle\sigma_z\rangle$ to the ideal final state to obtain the fidelity $\mathcal{F}$. There is an exponential decrease in $\mathcal{F}$ with an increasing number of gates in the randomized sequences. This RB protocol is then repeated for various pulse widths, corresponding to different Gaussian standard deviations, $\sigma \in \{1,2,3,4,6\}\,\ns$.

Using the Gaussian shaping, we find a large reduction in fidelity with the shortest pulses, $\sigma = 1\,\ns$ [Fig.\,\ref{fig:fig2}(a)]. The scattered grey points give $\mathcal{F}$ for 32 different randomized sequences applied as a function of the number of gates in the sequences. When averaged together, we observe a simple decay of $\bar{\mathcal{F}}$  as a function of the number of gates (solid black squares). Fitting the data with an exponential decay (solid black line), we extract an average error per gate, $\text{EPG} = 1- \bar{\mathcal{F}} $ of $0.13\pm 0.02$. However, when employing the first-order DRAG, we find a dramatic improvement in the gate performance at $\sigma=1\,\ns$ [Fig.\,\ref{fig:fig2}(b)]. There is a significant reduction in the spread of the grey points corresponding to all the different randomized sequences, and a fit (solid black line) to the exponential decay of the average fidelity (solid black squares) gives $\text{EPG}=0.007\pm 0.005$.

\begin{figure}[t!]
\centering
\includegraphics[scale=1.0]{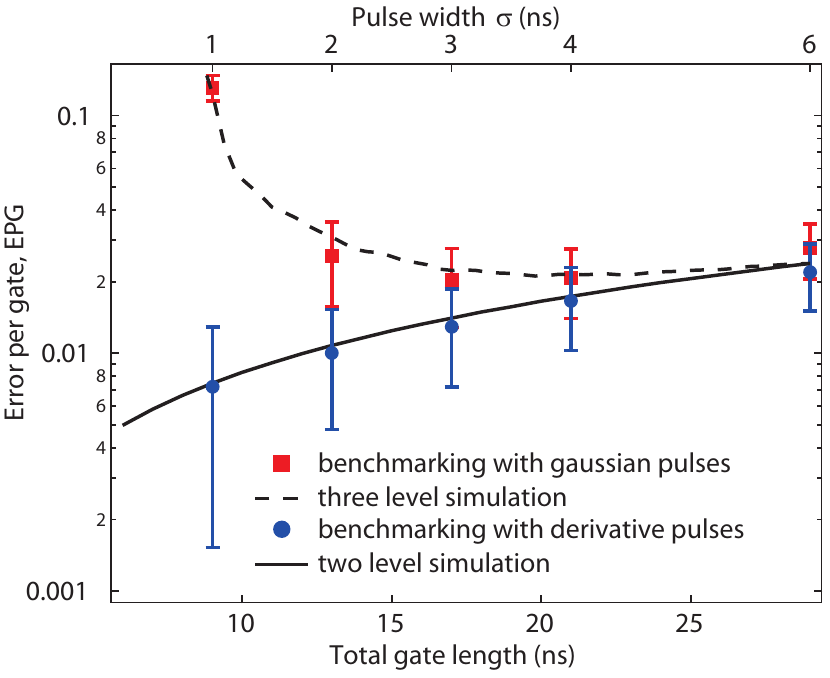}
\caption{(color online)~Comparison of single-qubit gate errors with and without DRAG. Error per gate for the left qubit extracted from randomized benchmarking for different gate lengths using both Gaussian pulses (red squares) and first-order DRAG pulses (blue squares). Excellent overlap with theory for gate error including qubit decoherence (black curve) suggests that the DRAG pulses
successfully eliminate the errors due to the presence of higher levels. Gate errors down to $0.007$, which are otherwise unattainable with Gaussian pulses, are reached using DRAG.}
\label{fig:fig3}
\end{figure}

Figure\,\ref{fig:fig3} summarizes the improvement to EPG for different $\sigma$ by using DRAG. The solid squares are the EPG found using Gaussians, revealing a minimum of $0.02\pm 0.007$ at $\sigma=3\,\ns$, before considerably increasing for shorter pulse lengths. Excellent agreement is found with a master equation simulation (dashed line) of the gate error for a qutrit system incorporating only decoherence times and coupling strengths measured in independent experiments. Using first-order DRAG, we find the solid circles in Fig.\,\ref{fig:fig3}, which follow a monotonic decrease in $\text{EPG}$ with decreasing $\sigma$. Here again we have included a master equation prediction (solid line) of just a single qubit with the same parameters, also giving excellent agreement with the experimentally determined values and demonstrating that DRAG has reduced the response of the system to be like that of a single qubit.

\begin{figure}[bpt!]
\centering
\includegraphics[scale=1.0]{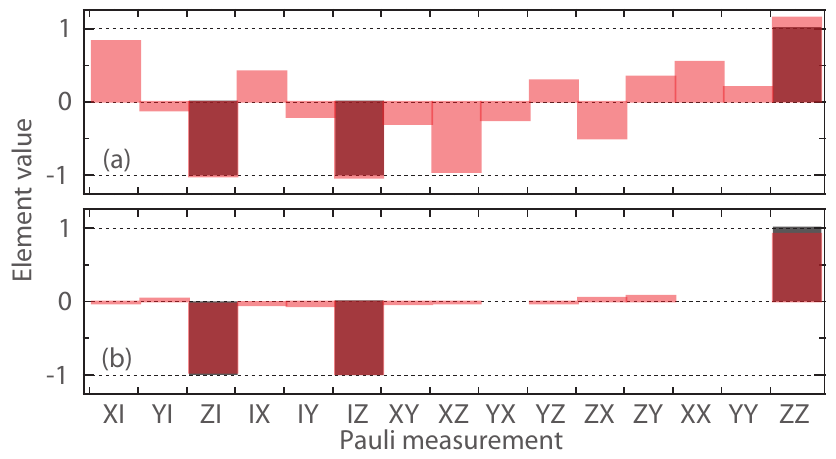}
\caption{(color online)~Measured two-qubit Pauli sets for preparing the state $\ket{1,1}$ with (a)~ Gaussian pulses and (b)~DRAG pulses ($\beta^{\text{L}}=0.4$, $\beta^{\text{R}}=0.25$) applied to the quadrature channels of both transmon L and qubit R. The ideal Pauli set is shown in grey.}
\label{fig:fig4}
\end{figure}

Finally, implementing DRAG on both qubits simultaneously, we can also generate and detect higher quality two-qubit states. Performing state tomography to obtain the two-qubit density matrix $\rho$ via joint readout \cite{filipp_joint_2009,chow_entang_2009} requires 15 linearly independent measurements, corresponding to the application of all combinations of $I$, $R_x^{\pi}$, $R_x^{\pi/2}$, and $R_y^{\pi/2}$ on the two qubits prior to measurement. Thus, errors in these analysis rotations in addition to the state preparation pulses can result in incorrect determination of $\rho$. The two-qubit Pauli set $\vec{P}$ \cite{chow_entang_2009} can be used to visualize $\rho$ for the state $\ket{1}_{\text{L}}\otimes\ket{1}_{\text{R}}$ having used Gaussian [Fig.\,\ref{fig:fig4}(a)] and DRAG [Fig.\,\ref{fig:fig4}(b)] pulse shaping. $\vec{P}$ consists of ensemble averages of the $15$ non-trivial combinations of Pauli operators on both qubits.  The ideal $\vec{P}$ of the state is characterized by unit magnitude in $\langle ZI \rangle$, $\langle IZ \rangle$, and $\langle ZZ\rangle$ and zero for all other elements. We can see that with the standard Gaussian pulse shaping, there are substantial ($\sim 50-100\%$ of unity) deviations on ideally zero elements, whereas with the DRAG pulses, the Pauli set bars are very close to their ideal values. 

By implementing a simple approximation to the optimal control pulses for a multi-transmon coupled-cavity system, we have reduced gate errors below the $10^{-2}$ level, limited by decoherence. The agreement of the various experiments with and without DRAG pulse shaping with a qutrit model reflects that gate errors due to the coupling to a higher excited state can be minimized while continuing to shorten gate time. Moving forward with optimal control, a tenfold decrease in gate time to approach $\sim 1\,\ns$ through improved electronics or a tenfold increase in coherence times to $\sim 10\,\us$ would place us right at the quoted $10^{-3}$ fault-tolerant threshold \cite{Motzoi:2009ca}. Furthermore, DRAG is extendable to systems of more than two multi-level atoms for quantum information processing, and has already been employed to enhance single-qubit gates in a circuit QED device with four superconducting qubits~\cite{dicarlo_GHZ_2010}.

\begin{acknowledgments}
	We acknowledge discussions with Erik Lucero, Lev S. Bishop, and Blake R. Johnson. This work was supported by LPS/NSA under ARO Contract No.\,W911NF-05-1-0365,
and by the NSF under Grants No.\,DMR-0653377 and No.\,DMR-0603369.
We acknowledge additional support from a CIFAR Junior Fellowship, MITACS, MRI, and NSERC (JMG), and from CNR-Istituto di Cibernetica, Pozzuoli, Italy (LF).
\end{acknowledgments}

\end{document}